\documentclass[final,3p,
twocolumn]{elsarticle}

\usepackage{times}
\usepackage{amsmath}
\usepackage{amsfonts}
\usepackage{amssymb}
\usepackage{bm}
\usepackage{txfonts}
\usepackage{ulem}
\usepackage{braket}
\usepackage[usenames]{xcolor}

\def\jcite #1#2#3#4{#1 {\bf #2}, #3 (#4).}

\journal{Physica B}

\begin{document}
\begin{frontmatter}
\title{Electronic band structure of 4$d$ and 5$d$ transition metal trichalcogenides}
\author[label1]{Yusuke Sugita}
\author[label2]{Takashi Miyake}
\author[label1]{and Yukitoshi Motome}
\address[label1]{Department of Applied Physics, University of Tokyo, Bunkyo, Tokyo 113-8656, Japan}
\address[label2]{CD-FMat, National Institute of Advanced Industrial Science and Technology (AIST), Tsukuba, Ibaraki 305-8568, Japan}

\begin{abstract}
Transition metal trichalcogenides (TMTs), a family of  van der Waals materials, have gained increasing interests from the discovery of magnetism in few-layer forms.
Although TMTs with 3$d$ transition metal elements have been studied extensively, much less is explored for the 4$d$ and 5$d$ cases, where the interesting interplay between electron correlations and the relativistic spin-orbit coupling is expected.
Using {\it ab initio} calculations, we here investigate the electronic property of TMTs with 4$d$ and 5$d$ transition metal elements.
We show that the band structures exhibit multiple node-like features near the Fermi level. 
These are the remnant of multiple Dirac cones that were recently discovered in the monolayer cases. 
Our results indicate that the peculiar two-dimensional multiple Dirac cones are concealed even in the layered bulk systems.  
\end{abstract}

\end{frontmatter}

\section{Introduction} 
Magnetism in van der Waals (vdW) materials has drawn considerable attention owing to the possibility of two-dimensional magnetism in the quasi-two-dimensional structures. 
Amongst others transition metal trichalcogenides (TMTs), which have a layered structure of the honeycomb network of TM cations, have intensively studied as a promising candidate. 
Indeed, a variety of magnetic states were discovered in the bulk form of 3$d$ TMTs, such as ferromagnetic, N\'{e}el, and zigzag antiferromagnetic states~\cite{BREC19863,OUVRARD198827,carteaux1995crystallographic}, and very recently, even in the atomically-thin form~\cite{doi:10.1021/acs.nanolett.6b03052,2053-1583-3-3-031009,Gong2017}.
From the theoretical point of view, previous {\it ab initio} studies suggested the importance of strong electron correlations of 3$d$ electrons in the TMTs~\cite{kurita1989,kurita1989band,zhukov1996electronic}, and such studies have been extended to the monolayer form~\cite{PhysRevB.91.235425,PhysRevB.94.184428}.  

Besides the magnetism, TMTs have also attracted interest due to the peculiar properties arising from the relativistic spin-orbit coupling (SOC). 
Recent theoretical works predicted anomalous electronic and transport properties induced by the SOC, such as valley splitting of the band structure~\cite{Li05032013}, the spin Nernst effect~\cite{PhysRevLett.117.217202}, and the magneto-optic Kerr effect~\cite{PhysRevLett.117.267203}.
Although the SOC effect will become more conspicuous for 4$d$ and 5$d$ elements than 3$d$, there are less studies focusing on 4$d$ and 5$d$ TMTs. 
Recently, the authors studied the electronic and magnetic properties of monolayer TMTs with 4$d$ and 5$d$ TMs, and showed that the compounds have a peculiar electronic band structure with multiple Dirac cones, which potentially turns into a Chern insulator by the interplay between the SOC and electron correlations~\cite{sugita2017multiple}.
However, as the study was limited to the monolayer form, it remains unclear how such interesting features are compiled in the bulk before the two-dimensional exfoliation.   

In this paper, we explore the electronic band structures of the 4$d$ and 5$d$ TMTs in the bulk form.
We show by {\it ab initio} band calculations that the electronic states have multiple node-like structures near the Fermi level, independent of the wave vector perpendicular to the honeycomb planes. 
These structures look similar to the multiple Dirac cones found in the monolayer systems.
Thus, our results indicate that the multiple Dirac cones are already hidden in the bulk systems.

This paper is organized as follows.
In Sec.~\ref{detail}, we introduce the crystal structure of TMTs and the {\it ab initio} method used in this paper.
In Sec.~\ref{result}, we show the band structures in the three-dimensional Brillouin zone and on the two-dimensional slices.
Section~\ref{summary} is devoted to the summary.

\section{{\it Ab initio} calculations}
\label{detail}
\begin{figure}[t]
\centering
\includegraphics[width=1.0\columnwidth,clip]{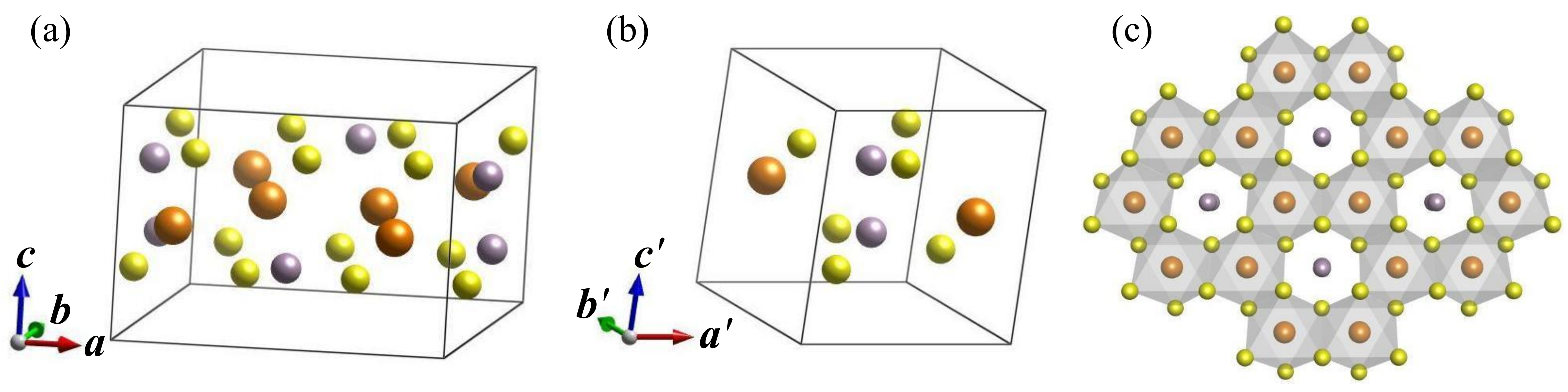}
\caption{
(a) and (b)~Monoclinic crystal structures of group 10 TMTs $M$PS$_3$ in the conventional cell and primitive cell, respectively~\cite{ZAAC:ZAAC19733960305}.
The orange, purple, and yellow spheres denote $M$, P, and S, respectively.
The translational vectors are denoted by $\bm{a}$, $\bm{b}$, and $\bm{c}$ in (a), while $\bm{a}'=(\bm{a}+\bm{b})/2$, $\bm{b}'=(-\bm{a}+\bm{b})/2$, and $\bm{c}'=\bm{c}$ in (b); we take $\bm{a}$ and $\bm{b}$ in the $xy$ plane.
(c)~Top view of a layer of TMTs.
The gray octahedra indicate the edge-sharing $M$S$_6$.
}
\label{setup}
\end{figure}

The general chemical formula of TMTs is given by $M$$B$$X_3$, where $M$ is a TM, $B$=P, Si, or Ge, and $X$ is a chalcogen~\cite{BREC19863,OUVRARD198827,carteaux1995crystallographic,ZAAC:ZAAC19733960305}.
The syntheses of TMT with 4$d$ and 5$d$ elements were reported for a group 10 element Pd and group 12 elements Cd and Hg~\cite{ZAAC:ZAAC19733960305}.
The TMTs with group 10 TM elements, $M$PS$_3$ with $M=$ Ni and Pd, take a monoclinic structure (the space group $C2/m$)~\cite{ZAAC:ZAAC19733960305}, as shown in Figs.~\ref{setup}(a) and \ref{setup}(b).
The lattice structure is quasi-two-dimensional and composed of atomic layers, where $M$S$_6$ octahedra form an edge-sharing honeycomb network and P$_2$ dimers locate at the centers of hexagons [see Fig.~\ref{setup}(c)]. 

In the following, we calculate the electronic band structure for the paramagnetic state of $M$PS$_3$ with $M=$ Pd (4$d$) and Pt (5$d$) with the monoclinic structure, by using OpenMX code~\cite{openmx}. 
We adopted the local density approximation (LDA) with the Perdew-Zunger parametrization for the exchange-correlation functional~\cite{PhysRevB.23.5048} in the density functional theory and a $8\times8\times8$ $\bm{k}$-point mesh for the calculations of the self-consistent electron density and the structural relaxation.
We optimized the lattice structure starting from the reported crystalline data for PdPS$_{3}$~\cite{ZAAC:ZAAC19733960305} by relaxing the primitive translational vectors and atomic positions in a non-relativistic {\it ab initio} calculation with the convergence criterion less than 0.01~eV/\AA~about the inter-atomic forces.
Using the optimized structures, we calculated the electronic band structures by a relativistic {\it ab initio} calculation, where the relativistic effects are included by a fully relativistic $j$-dependent pseudopotential.

\section{Result}
\label{result}
\begin{figure}[t]
\centering
\includegraphics[width=1.0\columnwidth,clip]{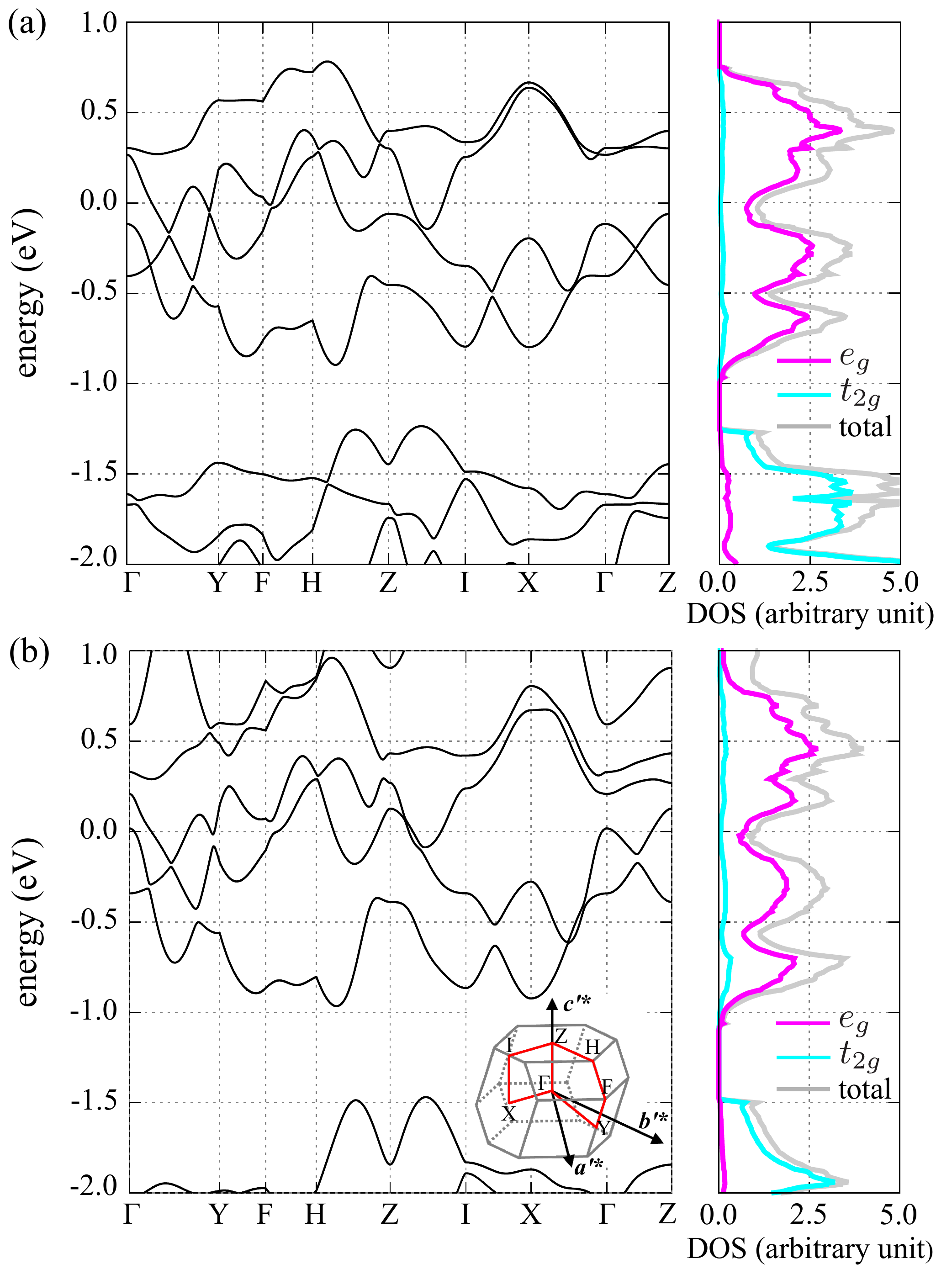}
\caption{
(a) and (b)~Electronic band structures and DOS for PdPS$_3$ and PtPS$_3$, respectively.
The data are plotted along the symmetric lines in the first Brillouin zone for the monoclinic crystalline structure of TMTs~\cite{SETYAWAN2010299} as shown in the inset of (b). 
Here, $\bm{a}'^{*}$, $\bm{b}'^{*}$, and $\bm{c}'^{*}$ are the reciprocal vectors corresponding to the primitive vectors $\bm{a}'$, $\bm{b}'$, and $\bm{c}'$ in Fig.~\ref{setup}(b).
The Fermi level is set to be zero. 
The total DOS (gray) is decomposed into the contributions from the $e_g$ and $t_{2g}$ electrons in Pd and Pt. 
}
\label{band}
\end{figure}

Figure~\ref{band} shows the band structures and the density of states (DOS) for PdPS$_3$ and PtPS$_3$ in the paramagnetic state.
In each case, the electronic states near the Fermi level are dominated by $e_{g}$ orbitals of $d$ electrons.
This is because the nominal valence of TMs is given by $M^{2+}$, and thus Pd and Pt are in the $d^8$ configuration, for which the $e_g$ band is half filled.
Although the $e_g$ bandwidth is broader for PtPS$_3$ than PdPS$_3$, the band structures have a similar structure. 
In particular, there exist several band crossings near the Fermi level along the $\Gamma$-Y-F-H-Z lines, which appear to lead to a dip structure in the DOS.
We note that the previous {\it ab initio} study for the monolayer form reported the multiple Dirac cones near the Fermi level along the $\Gamma$-K line~\cite{sugita2017multiple}.
The band crossings in Fig.~\ref{band}, therefore, may be related to the two-dimensional multiple Dirac cones.  
Indeed, the $\Gamma$-Y-F-H-Z lines are on the plane spanned by $\bm{a}'^{*} + \bm{b}'^{*}$ and $\bm{c}'^{*}$ [see the inset in Fig.~\ref{band}(b)], which suggests that they correspond to the $\Gamma$-K line by the projection onto the two-dimensional Brillouin zone for the monolayer system.

\begin{figure}[t]
\centering
\includegraphics[width=1.0\columnwidth,clip]{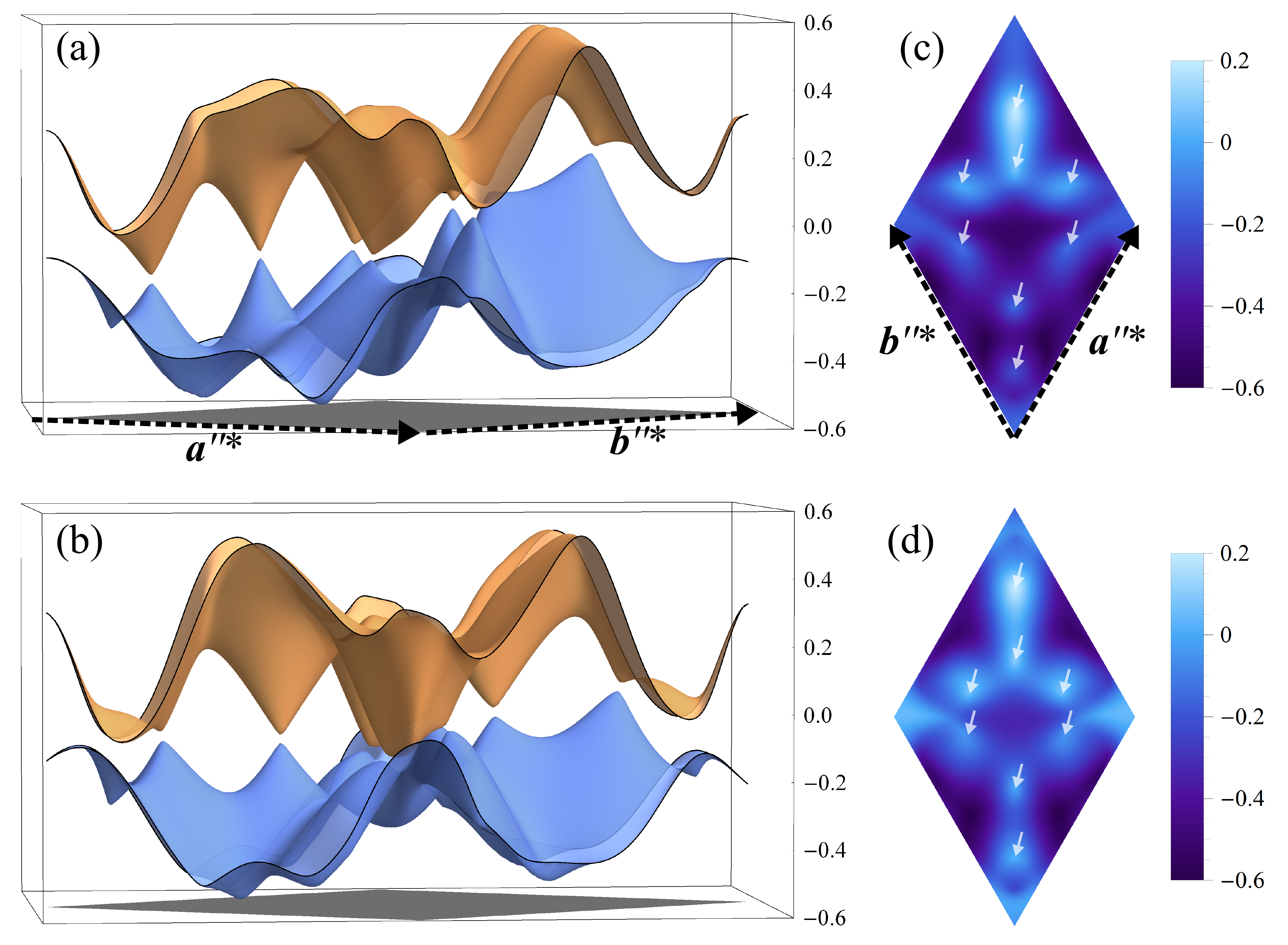}
\caption{
(a) and (b)~3D plots of two electronic bands near the Fermi level for PdPS$_3$ in the plane with $k_z =0$ and $k_z = \pi$, respectively.
$\bm{a}''^{*}$ and $\bm{b}''^{*}$ span the two-dimensional slices of the Brillouin zone parallel to the honeycomb layers; see the text for details. 
(c) and (d)~Contour plots of the low-energy bands (blue) bands in Figs.~\ref{splot}(a) and \ref{splot}(b), respectively.
The positions of Dirac nodes are denoted by the white arrows. 
}
\label{splot}
\end{figure}

By carefully looking into the band structures near the Fermi level, we confirm the relation between the band crossings and the multiple Dirac cones.
Figure~\ref{splot} shows the 3D plots of the two bands near the Fermi level for PdPS$_3$: (a) on the $k_z=0$ plane and (b) on the $k_z = \pi$ plane; $k_z$ is the wave number along the $z$ axis, and $\bm{a}''^{*}$ and $\bm{b}''^{*}$ are the reciprocal vectors spanning the two-dimensional slices of the Brillouin zone parallel to the honeycomb layers: $\bm{a}''^{*}=2\pi \left( \bm{b}' \times \bm{e}_z \right) /\left\{ \bm{a}' \cdot \left( \bm{b}' \times \bm{e}_z \right) \right\}$ and $\bm{b}''^{*}=2\pi \left( \bm{a}' \times \bm{e}_z \right) /\left\{ \bm{b}' \cdot \left( \bm{a}' \times \bm{e}_z \right) \right\}$ ($\bm{e}_z$ is the unit vector along $z$ axis) [see Fig.~\ref{setup}(b)].
The overall landscapes of the energy spectrum are qualitatively similar between $k_z=0$ and $\pi$, and we can see eight node-like structures. 
To show them more clearly, we present the energy contour plots of the lower bands in Figs.~\ref{splot}(c) and \ref{splot}(d).  
In both plots, there are eight white spots corresponding to the band crossings, as indicated by the white arrows.  
We find that the positions of the nodes in each two-dimensional slice of the Brillouin zone coincide well with those of the multiple Dirac cones in the monolayer cases~\cite{sugita2017multiple}.
We note that the node-like structures in the bulk case break the six-fold symmetry due to the monoclinic lattice structure, whereas the monolayer case preserves the symmetry~\cite{sugita2017multiple}.

\section{Summary}
\label{summary}
We have investigated the bulk electronic states of the 4$d$ and 5$d$ TMTs, for which the multiple Dirac cones in the band structures were theoretically discovered in the monolayer form~\cite{sugita2017multiple}.
We clarified that the TMTs have metallic bands in the paramagnetic states, whose low-energy states are dominated of the half-filled $e_{g}$ bands.
In addition, by showing the band structures projected onto the two-dimensional slices of the Brillouin zone, we elucidated that the bulk TMTs have similar node structures to the multiple Dirac cones found in the monolayer case.

There remain several open issues for the 4$d$ and 5$d$ TMTs.
For instance, it is an interesting challenge to investigate the influences of the layer number and stacking order on electronic states, which plays an important role in other vdW materials, e.g., graphene~\cite{RevModPhys.81.109}.
The edge states in few-layer forms and nano-ribbon shapes, which have also been studied in graphene~\cite{RevModPhys.81.109}, would be interesting.  
Furthermore, the effect of electron correlations is of great interest.

\section*{Acknowledgments}
Y.S. is supported by the Japan Society for the Promotion of Science through a research fellowship for young scientists and the Program for Leading Graduate Schools (MERIT).
The crystal structures are visualized by using VESTA 3~\cite{Momma:db5098}.

\end{document}